\documentclass{cplslarge}

\usepackage[authoryear]{natbib}
\usepackage{epsf}
\usepackage{graphicx}
\usepackage{amsmath,amssymb,aas_macros,url}
\begin{document}
% \frontmatter
% \maketitle
% \tableofcontents
% \maketitle
% \listoffigures
% \listoftables
\listofcontributors
\mainmatter
\chapter{\label{Chapter5}Extra-Terrestrial Meteors}
% \linenumbers
\section{\label{sec:intro}Introduction}
The beginning of the space age 60 years ago brought about a new era of discovery for the science of astronomy. Instruments could now be placed above the atmosphere, allowing access to new regions of the electromagnetic spectrum and unprecedented angular and spatial resolution.  But the impact of spaceflight was nowhere as important as in planetary and space science, where it now became possible -- and this is still the case, uniquely among astronomical disciplines -- to physically touch, sniff and directly sample the bodies and particles of the solar system. A cursory reading of the chapters in this volume will show that meteor astronomy has progressed in leaps and bounds since the era of visual observations. Indeed, it has evolved into a true analytical {\sl science}, on a par with sister disciplines such as cometary science and meteoritics. Unfortunately, instrumentation specifically designed to detect meteors on planets other than the Earth has yet to fly on a planetary mission. 
 
Two factors have now brought the era of exo-meteor astronomy closer than ever before. One is the serendipitous detection of meteors and their effects in the relatively thick atmospheres of Mars and Jupiter as well as the tenuous exospheres\index{surface boundary exosphere} of Mercury and the Moon, developments of sufficient import to warrant dedicated sections in this chapter.  The other is technological advances in the detection of short-lived luminous phenomena in planetary atmospheres and a new understanding of the unique operational aspects of off-Earth meteor surveys.       
Throughout this chapter we use the term ``extraterrestrial meteor'' or ``exo-meteor''. We base our use of this terminology on the distinction between meteoroids and meteors. While ``meteoroid'' refers to particulate matter in interplanetary space, a ``meteor''  is a {\it phenomenon} -- the light emitted during the entry of a meteoroid into an atmosphere -- that is intrinsic to that atmosphere. Therefore, the term ``extraterrestrial meteor'' as previously used in the literature \citep{Selsis.et.al2005} and the principal subject matter of this chapter, uniquely defines meteors in planetary atmospheres other than the Earth's.

Why study these meteors? Despite the current rapid progress of Earth-based meteor astronomy, our ability to do useful science will always be limited by the simple fact that we can only sample meteoroid populations with orbits that cross the Earth's. Yet, there is every reason to expect, for instance, that the solar system is criss-crossed with a web of dust streams\index{meteoroid streams} spread along cometary orbits, all-but-invisible except where the particles are packed densely enough to be detectable, typically near the comet itself \citep{SykesWalker1992,Reach.et.al2000,Gehrz.et.al2006}. Extending meteor observations to other planetary bodies allows us to map out these streams\index{meteoroid streams} and investigate the nature of comets whose meteoroid streams\index{meteoroid streams} do not intersect the Earth. Observations of showers corresponding to the same stream at two or more planets will allow to study a stream's cross-section. In addition, the models used to extract meteoroid parameters from the meteor data are fine-tuned, to a certain degree, for Earth's atmosphere. Processing exo-meteor data with these same models will be particularly useful in demonstrating their robustness or uncovering as-yet-unknown limitations that will impact our Earth-centred understanding of the meteor phenomenon. Last but not least, observations of exo-meteors will promote the safety of deep space missions, both crewed and robotic, by identifying regions where the meteoroid flux is high enough to pose significant risk to people or machines.

It is fortuitous that the atmospheres of both our neighbouring planets lend themselves to meteor observations. Mars combines a predominantly clear atmosphere\index{atmosphere, Mars} and a solid surface to serve as a stable observation platform. It is also the target of a vigorous international exploration program. Venus may, at first glance, not appear as appealing; apart from the extreme environmental conditions, the Venusian sky is perpetually hidden from the surface by planet-encircling cloud layers \citep{Esposito.et.al1983}. Yet this still leaves the possibility of observing meteors from above the atmosphere\index{atmosphere, Venus}, a technique already employed in observing meteor showers at the Earth \citep{Jenniskens.et.al2000}. 

\section{\label{sec:theory}Theoretical Expectations}
\begin{figure}[!htbp]
\figurebox{30pc}{}{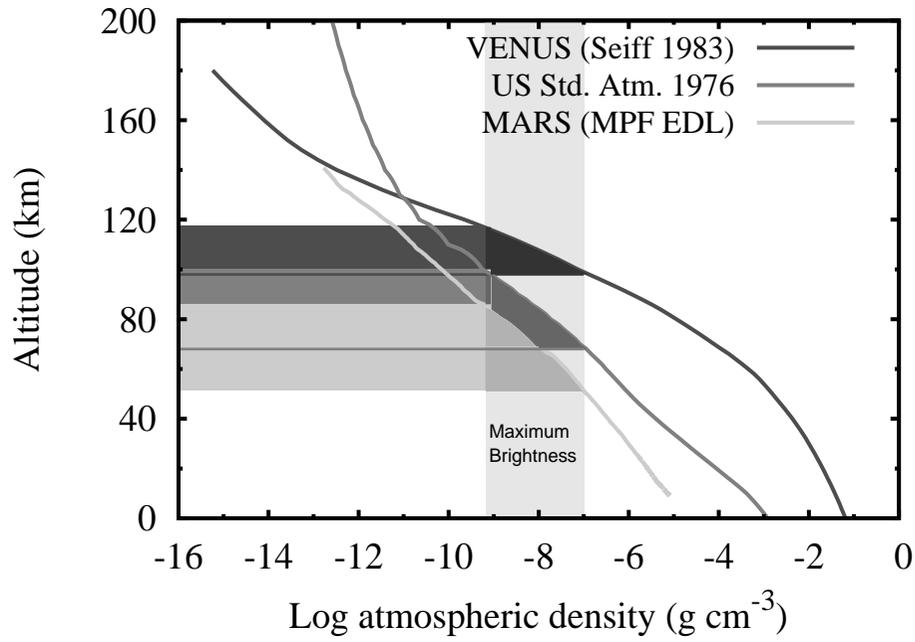}
\caption{Atmospheric density-height profiles for different planets. From top to bottom: Venus (“nightside” atmospheric model by \citet{Seiff1983}  -- dark grey line), Earth (US Standard Atmosphere 1976 -- moderate grey line) and Mars ({\it Mars Pathfinder} entry profile -- light grey line). The horizontal bands indicate the respective height ranges -- corresponding to the same density range -- at which meteors reach maximum luminosity. Adapted from \citet{Christou2004a}.}
\label{fig:atmo}
\end{figure}

Single body meteoroid ablation\index{meteoroid ablation} in the Martian atmosphere\index{atmosphere, Mars} has been modelled by several wofor erks \citep{Apshtein.et.al1982,Adolfsson.et.al1996,McAuliffe2006}. 
\citet{Adolfsson.et.al1996} compared the height and brightness of meteors at Mars and the Earth, assuming that these properties do not sensitively depend on atmospheric composition. They found that, in general, meteors reach their maximum brightness within the same range of atmospheric density (Figure~\ref{fig:atmo}). Martian meteors ablate between altitudes of 90 and 50 km; fast (30 km\,s$^{-1}$), low-density (0.3 g\,cm$^{-3}$) meteoroids in the two atmospheres generate meteors of similar brightness while slower, denser (3 g\,cm$^{-3}$) particles produce significantly fainter meteors at Mars. They were able to reproduce their numerical results with the intensity law
\begin{equation}
I_{m} \propto m \mbox{\textsl{\textrm{v}}}^{3+n}/H
\end{equation}
where $m$ is the meteoroid's mass, \textsl{\textrm{v}} its atmospheric impact speed, $H$ the density scale height and $n$ depends on the meteoroid type as described above. \citet{Christou2004a} applied the same principles to the case of Venus, finding that, everything else being equal, Venusian meteors would be as bright or brighter than at the Earth but also shorter-lived. Using the same argument as \citeauthor{Adolfsson.et.al1996}, whereby the meteoroid ablation\index{meteoroid ablation} rate peaks between 10$^{-9.2}$ and 10$^{-7}$ g\,cm$^{-3}$, meteors would appear between 100 and 120 km in the venusian atmosphere\index{atmosphere, Venus} and above the haze layer \citep{Esposito.et.al1983}, boding well for meteor searches from Venus orbit. \citet{McAuliffeChristou2006} confirmed these earlier conclusions with a numerical ablation\index{meteoroid ablation} model that takes into account radiative surface cooling and heat conduction within the meteoroid, effects not considered in \citet{Christou2004a}. Additional numerical work has been carried out in the context of the presence of layers\index{metal layers} of meteoric metals -- in both neutral and ionic form -- in the upper atmospheres of these planets with emphasis on Mars \cite[eg][see Section~\ref{sec:iono}]{PesnellGrebowsky2000}. A first step for such studies has been to determine the mass deposition rate of individual atomic metal species (Fe\index{Fe, neutral}, Mg\index{Mg, neutral}) from the ablating meteoroid as a function of altitude. The deposition rates of these species represent different components of the gross ablation\index{meteoroid ablation} rate of the meteoroid, equivalent to meteor brightness in the single body ablation\index{meteoroid ablation} model. The results indicate -- as in the earlier works -- that the maximum brightness of martian meteors is reached $\sim$10 km lower in the atmosphere\index{atmosphere, Mars} and over a similar or somewhat broader range of altitudes compared to terrestrial meteors. Similar calculations were done as part of a recent study \citep{Frankland.et.al2017} of O$_{2}$ removal efficiency at Venus by CO$_{2}$ oxidation on meteoric remnant particles. These show, as for Mars, that the maximum deposition rate of meteoric metals agrees with the earlier predictions from single body ablation\index{meteoroid ablation} modelling, but also indicate that significant ablation\index{meteoroid ablation} may be taking place at altitudes as high as 130--135 km.   

\section{\label{sec:obs}Observational Record}
As mentioned in the Introduction, no direct optical detection of exo-meteors has been achieved to-date. A transient event observed by the Optical and Ultraviolet Visible Spectrometer ({\it OUVS}) instrument onboard the {\it Pioneer Venus Orbiter\index{Pioneer Venus Orbiter spacecraft}} on 1979 February 17 \citep{huestis1993} was interpreted as a serendipitous detection of a meteor trail but remains unconfirmed.  A detection of a meteor associated with a known Jupiter-family comet by the dual-eye Pancam imager onboard the {\it Spirit\index{Mars Exploration Rovers}} rover on Mars was reported by \citet{Selsis.et.al2005}. Later work by \citet{Domokos.et.al2007} showed that the claimed 2005 detection was most likely a cosmic ray hit on the Pancam detector.  In addition, the non-availability of a broadband or clear filter on both Pancam eyes severely limited the scope of the search, which did not yield any definite detections. This highlights the need for dedicated instruments to carry out exo-meteor observations.

Although not a search for exo-meteors {\sl per se}, it is worth mentioning here a successful recording of a meteor shower from space. The $13^{\circ}$$\times$$10^{\circ}$ Wide Field Imager (WFI) channel of the Ultraviolet and Visible Imagers and Spectrographic Imagers (UVISI) suite on board the {\it Midcourse Space eXperiment\index{Midcourse Space eXperiment spacecraft}} (MSX) satellite logged 29 meteor detections during an effective observing time of $\sim$20 min on 1997 November 17. The Leonid flux derived from these observations was (5.5$\pm$0.9)$\times$${10}^{-2}$ km$^{-2}$\,hr$^{-1}$ down to a limiting absolute magnitude of $-$$1.5^{\rm m}$ and a population index of r$=$1.7 \citep{Jenniskens.et.al2000}. It demonstrates that detection of even moderately strong meteor showers with instruments not specifically suited to the task is possible from orbit.
\section{\label{sec:pred}Predictions}
\subsection{\label{ssec:observ}Observability}
For Mars, \citet{Adolfsson.et.al1996} estimated the flux of meteors of absolute visual magnitude\footnote{The magnitude a visual observer would perceive of a meteor at 100 km distance and at the zenith} $-$1$^{\rm m}$ -- $+$4$^{\rm m}$ at Mars to be 50\% of that at the Earth. \citet{Domokos.et.al2007} placed an upper flux limit of 4.4 $\times$ $10^{-6}$ km$^{-2}$\,hr$^{-1}$ for $>$ 4g meteoroids, consistent with the \citeauthor{Adolfsson.et.al1996} prediction (4 $\times$ $10^{-7}$ km$^{-2}$\,hr$^{-1}$) derived by their scaling of the Earth flux according to the \citet{Grun.et.al1985} model. Such flux estimates, though approximate, may be used to estimate sporadic meteor detection rates for a given camera system from specific vantage points. For instance, the expected  detection rate by a meteor camera system specifically designed for operation in space \cite[Smart Panoramic Optical Sensor Head or SPOSH;][]{Oberst.et.al2011} is between 1 and 7 meteors per orbit or between 14 and 74 per Earth day ($\sim$11 nightside passes) for a 400 km altitude circular orbit at a limiting apparent magnitude of $+2^{\rm m}$ \citep{Christou.et.al2012}. \citet{Bouquet.et.al2014} find a similar detection rate for a SPOSH-like system operating in Earth orbit.

For Venus, \citet{BeechBrown1995} investigated the feasibility of observing bright fireballs (entry mass of $10^{4}$-$10^{7}$ kg) in the venusian atmosphere\index{atmosphere, Venus} from the Earth. They found, for instance, that events with apparent magnitude $+10^{\rm m}$ -- equivalent to an absolute magnitude of  $-24^{\rm m}$ -- or brighter will occur every three weeks whereas fainter, $+15^{\rm m}$ events of $-19^{\rm m}$ meteor absolute magnitude will occur every 2 days. The authors pointed out that the flux could be higher at times when streams\index{meteoroid streams} rich in fireball-producing meteoroids cross the orbit of Venus and advocated long-term monitoring programmes to determine whether such fireball-rich streams\index{meteoroid streams} exist. 
Interestingly, \citet{Hansell.et.al1995} reported the detection of \adjustfigure{2.5cm} seven flashes of (1 -- 20)$\times 10^{9}$ J in luminous energy from narrowband observations at 777 nm with an effective observing time of 3 hr, a rate of $2.7 \times 10^{-12}$ km$^{-2}$\,sec$^{-1}$ which they attributed to lightning. This is 6$\times$(2$\times$24h)/3h = 100 times the \citeauthor{BeechBrown1995} rate for $-20^{\rm m}$ fireballs assuming a $10^{-3}$ luminous efficiency. 
\subsection{\label{ssec:showers}Showers and Outbursts}
It has been known since ancient times that meteor activity is not uniformly random but tends to concentrate around specific times of the year. The first association between a comet and a meteor shower was made 150 years ago \citep{Schiaparelli1867} with the first systematic survey of meteor shower activity carried out in the middle of the 20th century \citep{Lovell1949,Aspinal.et.al1949}. Repeating the same exercise on other planets, we are faced with the added difficulty that there are essentially no observations, but a multitude of potential parent bodies that may be producing strong meteor activity or nothing at all. To illustrate the problem, consider the case of (3200) Phaethon\index{asteroid (3200) Phaethon} associated with one of the most prolific annual showers, the December Geminids \citep{Whipple1983}. The nature of this object has been debated for decades \citep{Ryabova2015,Ryabova2018}. Recent observations from the Sun-orbiting {\it 
} spacecraft indicate activity of an exceptional type, not related to release of dust grains from within a sublimating matrix \citep{LiJewitt2013}. Arguably, the motivation to study and explain \index{asteroid (3200) Phaethon} arose because of the ``happy coincidence'' whereby the Geminid stream intersects the orbit of the Earth.

The distance of the comet orbit to the planet orbit ($\Delta$\index{orbit-to-orbit distance}) offers a necessary, but not sufficient, criterion for a comet to produce an observable meteor shower. Because it is computationally expedient to calculate it for a large number of orbits, it has been employed time and again in the literature \citep{Terentjeva1993,ChristouBeurle1999,TreimanTreiman2000,Larson2001,Christou2004a,Selsis.et.al2004,Neslusan2005}. A byproduct of these works is that we now have a fairly strong grasp on the population characteristics of planet-approaching comets for Mars, Venus and the outer planets as compared to the Earth's. They allowed to identify the best candidate parent bodies as input to more sophisticated techniques.

The observational record at the Earth teaches us that the minimum approach \index{orbit-to-orbit distance} -- or its projection on the orbit plane -- cannot be the sole discriminator for meteor activity. While it works for e.g.~the Perseids in August (109P/Swift-Tuttle), Leonids in November\index{Leonids meteor shower} (55P/Tempel-Tuttle) and Lyrids in April (C/1861 G1 (Thatcher)) in the sense that $\Delta$ takes particularly small values for these objects, it fails in several prominent cases such as the Taurids (2P/Encke\index{comet 2P/Encke}), $\eta$ Aquariids and Orionids (1P/Halley\index{comet 1P/Halley}) where $\Delta$\index{orbit-to-orbit distance} is of order 0.1 au or greater. An important additional factor is the dynamical type of the orbit. Because of the efficient action of Jupiter in rapidly scattering the orbits of dust grains away from the comet's, annually recurring meteor activity from Jupiter Family Comets is typically weak or non-existent and large numbers of meteors appear only during outbursts e.g.~the Draconids in October (21P/Giacobini-Zinner) or the Bootids in June (73P/Schwassmann-Wachmann 3).

\begin{figure}
\figurebox{30pc}{}{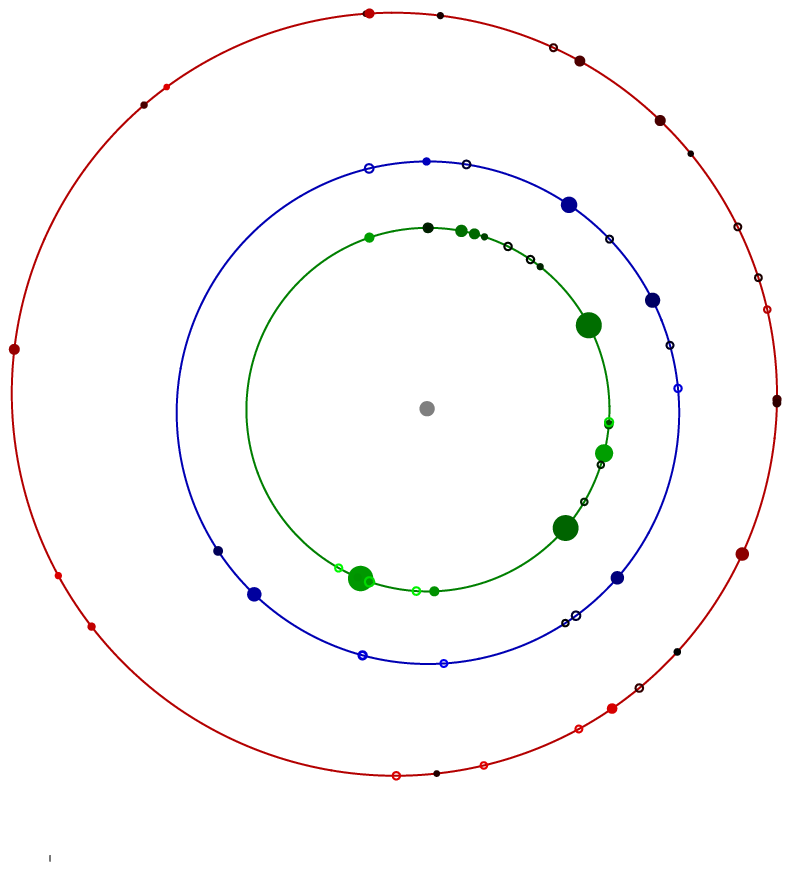}
\caption{Locations of meteoroid streams\index{meteoroid streams} encountering the orbits of Earth (blue), Venus (green) and Mars (brown) based on data from \citet{Christou2010}. The First Point of Aries is towards the right. Open symbols correspond to Encke-type Comets, filled symbols to Intermediate Long Period and Halley Type Comets. The size of each symbol is proportional to the encounter speed while the brightness indicates the solar elongation of the radiant.}
\label{fig:showers}
\end{figure}

To identify the best shower candidates, \citet{Christou2010} refined the $\Delta$-search technique by exploiting the tendency for strong meteor showers to be associated with Halley-type and Encke-type comets, rather than Jupiter-family and Long-period comets. The final distillation of the best candidates among potential parent bodies and their distribution along the planetary orbit is illustrated in Figure~\ref{fig:showers}. Note the relative lack of objects on the left half of each orbit, suggesting a semi-annual modulation of the shower frequency. This is likely a consequence of the method since the observed level of terrestrial meteor activity is a contributing factor in the predictions. Future surveys at Mars and Venus should confirm this.

Arguably, modern meteor forecasting was brought on by (a) the advent of cheap computing power, and (b) adopting the so-called {\sl trail} model of cometary dust evolution \citep{KondratevaReznikov1985,McNaughtAsher1999} where populations of meteoroids ejected from the comet at a given perihelion passage maintain their cohesiveness in space as distinct ``trails'' of particles over many orbital revolutions. The key result is that the dynamical evolution of trails is {\it deterministic}, so meteor outbursts can be reliably forecasted with numerical simulations of large number of test particles to serve as tracers of the dust evolution. The potential of this approach was spectacularly demonstrated during the Leonid meteor storms in 1999 and 2001. Guided by trail model predictions, observational campaigns are now routinely organised in advance to study debris reaching the Earth from the same comet at different years, from different comets as well as from different comet types \citep{Wiegert.et.al2011,Koten.et.al2015,Ye.et.al2015}. 
 
 It was only a matter of time until the trail method found application in exo-meteor shower prediction. It was used by \citet{VaubaillonChristou2006} and \citet{Christou.et.al2007} to identify dust trail encounters between Jupiter-Family comets 45P/Honda-Mrkos-Pajdusakova with Venus and 76P/West-Kohoutek-Ikemura with Mars respectively. \citet{Christou.et.al2008} simulated dust ejection from comet 1P/Halley\index{comet 1P/Halley} $\sim$5,000 yr ago to form a model of the stream and study how the characteristics of the corresponding shower differ from Venus to Earth and to Mars. 
 
\begin{figure}[!htbp]
\centering
\figurebox{30pc}{}{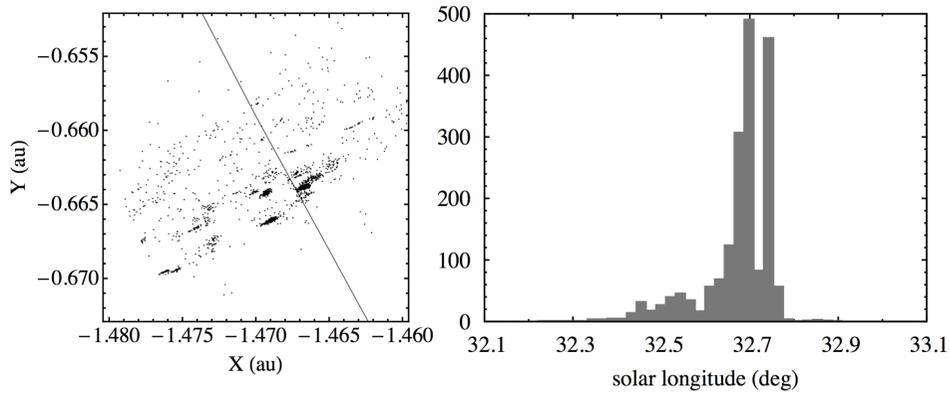}
\caption{{\sl Left}: Nodes of test particles from comet C/2007 H2 (Skiff)\index{comet C/2007 H2 (Skiff)} that encounter Mars between the years 2000 and 2050. The points represent cartesian heliocentric J2000 coordinates and units of au, as the particles cross the martian orbit plane. The black curve is the orbit of Mars, with the planet's direction of motion being from top to bottom. Note the numerous concentrations of particles within the stream's cross-section. {\sl Right}: Histogram of those particles on the left panel that approach the planet's orbit to within $0.005$ au as a function of solar longitude in degrees. Each bin corresponds to a time interval of one hour. After \citet{ChristouVaubaillon2011}.}
\label{fig:2007H2}
\end{figure}

The same approach was used by \citet{ChristouVaubaillon2011} on a study of larger scope, namely to simulate the meteoroid streams\index{meteoroid streams} of the parent body candidates in \citet{Christou2010}. They found that particles from many of these comets physically approach Mars and Venus and do so year after year, suggesting the presence of annually-recurring activity. The typical efficiency of particle delivery -- in other words the fraction of particles physically encountering a planet out of those ejected -- is $\sim10^{-4}$ per comet per planetary year.  For six of these, however, the model dust distribution cross-section appears highly inhomogeneous, capable of producing outbursts of activity well above the annual level. We show one such case, the stream of comet C/2007 H2 (Skiff)\index{comet C/2007 H2 (Skiff)}, in Figure~\ref{fig:2007H2}. The dust distribution is reminiscent of that of the Leonids\index{Leonids meteor shower} at the Earth and suggests that our planet is not alone in possessing annual showers that produce outbursts on certain years. Tables \ref{tab:futureMSMars} and \ref{tab:futureMSVenus} show forecasts for Mars and Venus for the coming years, in addition to those in the literature \cite[e.g.][]{Christou2010,ChristouVaubaillon2011}. Interestingly, the width of a meteoroid trail decreases when its heliocentric distance decreases, leading to higher Zenithal Hourly Rate\footnote{The number of shower meteors that a visual observer would see on a clear night during one hour with the shower radiant at the zenith.} ($ZHR$) and shorter duration showers at Venus than at Mars. However, as pointed out by \cite{Christou2004b}, since the number of Mars-crossing is greater than Venus-crossing comets, we expect more meteor showers at Mars than at Earth and Venus. These two factors combine and we observe that the number of showers at Venus is quite small, and the $ZHR$ low, simply because the odds to encounter the planet are lower for Venus than for Mars.

The most recent work to model the meteoroid influx from individual comets on a planetary body has focused on the very close approach of comet C/2013 A1 (Siding Spring)\index{comet C/2013 A1 (Siding Spring)} to Mars \citep{Vaubaillon.et.al2014a,Moorehead.et.al2014,Tricarico2014,Farnocchia2014,Vaubaillon.et.al2014b} and an enhancement of the release rate of metallic species in Mercury's tenuous exosphere\index{surface boundary exosphere} probably caused by particles from 2P/Encke\index{comet 2P/Encke} \citep{Killen2015,Christou2015}. 
These are described in more detail in Sections~\ref{ssec:comet} and \ref{sec:exospheres} respectively.

A question linked to the orbital simulations is that of the ensemble atmospheric properties of the hypothetical showers and their detectability. This problem admits to Monte Carlo simulations and has been investigated either parametrically, by varying pertinent detector properties such as field-of-view (FOV), limiting magnitude and orbital vs surface-based vantage point \citep{McAuliffe2006, McAuliffeChristou2006} or by using specific instruments as the baseline \citep{Christou.et.al2012}. To indicate what is possible, we quote here some results for the annual Leonid shower from Appendix C of \citeauthor{McAuliffe2006}. A zenith-pointed instrument with $60^{\circ}$ FOV and limiting magnitude of $+2^{\rm m}$ will detect 70 meteors per hr and 51 meteors per orbital revolution respectively from the Earth's surface and from Earth orbit. The respective figures for Mars are 25 and 15 while from Venus orbit, 215 detections per orbital revolution are expected. For orbital detection, a lesson that applies to all these planets is the crucial role of the shower population index ie the slope of the meteoroid magnitude distribution and the speed. For a given Earth {\it ZHR}, fast showers with shallow magnitude distribution yield the highest number of detections. In this case, a large field-of-view is preferable to high sensitivity on account of the numerous bright meteors occurring far from nadir, a finding also true for sporadics \citep{Bouquet.et.al2014}. Surface-based meteor cameras at Mars will have to contend with the varying amount of atmospheric dust. While they may achieve several tens to a few hundreds of detections per hr at times when the atmosphere\index{atmosphere, Mars} is clear (dust optical depth $\tau=0.5$), during periods of high dust loading ($\tau=3.0$) the detection rate drops to a few per hr at best.  

In conclusion, work done so far suggests that (a) with the exception of a surface camera at Venus, camera systems specifically designed for meteor work will be no less useful at Venus and Mars than at the Earth, and that (b) one should expect comparable detection rates for orbital and surface-based meteor surveys.

\section{\label{sec:iono}Ionospheric Layers\index{metal layers} from Meteor Ablation\index{meteoroid ablation}}
\subsection{\label{ssec:overview}Overview and State-Of-The-Art}
Meteoroid ablation\index{meteoroid ablation} deposits all species common in meteoroids, such as O, Na\index{Na, neutral}, Mg\index{Mg, neutral}, Al, Si, K, Ca\index{Ca, neutral}, Ti, and Fe\index{Fe, neutral}, into the upper atmosphere \citep{Plane2018}. Since oxygen is already present in a typical atmosphere, trace amounts of meteoroid-derived oxygen have little effect. The metal species, however, are exotic constituents of atmospheres. They behave differently from the major constituents and trace amounts of metal species can have noticeable effects. These metallic species are deposited at altitudes where the neutral atmosphere is already weakly ionised by sunlight's ultraviolet photons. They can themselves be ionised by sunlight, by chemical reactions with existing atmospheric ions, or directly during ablation\index{meteoroid ablation}. Metallic species tend to form atomic ions like Mg$^{+}$\index{Mg, ions} and Fe$^{+}$\index{Fe,ions}, whereas the ambient ionosphere at ablation\index{meteoroid ablation} altitude is dominated by molecular ions \citep{schunk2009}. They are destroyed by transport downwards into denser regions of the atmosphere where atomic metal ions undergo three-body reactions to form molecular ions that are quickly neutralized by dissociative recombination with an electron \citep{molinacuberos2008}.

% \adjustfigure{105pt}
Since atomic ions cannot dissociatively recombine like molecular ions, atomic metal ions tend to be long-lived and a slow production rate of atomic ions can maintain a significant plasma population. 
Consequently, meteoroids affect the structure, chemistry, dynamics, and energetics of Earth's ionosphere. Compositional profiles have been obtained by 50 (as of 2002) sub-orbital rocket flights with in situ mass spectrometers \citep{grebowsky2002b}. They consistently show a metal ion layer a few km wide located between 90 km and 100 km. Observed properties are highly variable. The many ground-based ionosondes in continuous operation regularly detect post-sunset narrow plasma layers\index{metal layers} that contain long-lived atomic metal ions at the same altitudes. These are called Sporadic E layers. Models of terrestrial metal ion layers\index{metal layers} rely on wind shear in a strong and inclined magnetic field to organize ions into narrow layers\index{metal layers} \citep[e.g.,][]{carter1999}.

Do meteoroids also produce similar layers\index{metal layers} of metal ions in the ionospheres\index{ionospheres, planetary} of other solar system objects? 
The basic input exists throughout the solar system as meteoroids enter all solar system atmospheres at orbital speeds. However, ionospheric chemistry and magnetically-controlled plasma dynamics will differ between Earth and other objects \citep{molinacuberos2008, schunk2009}.
Theoretical models of the effects of meteoroids have been developed for many solar system ionospheres\index{ionospheres, planetary}
\citep[e.g.,][and references therein]{lyons1995,pesnell2000,kim2001,molinacuberos2003,moses2000,whalley2010,molinacuberos2008}. In general, these models predict that the ablation\index{meteoroid ablation} of meteoroids should produce a narrow plasma layer a few neutral scale heights below the main ionospheric peak. For simplicity, these models generally focus on Mg\index{Mg, ions} and Fe\index{Fe, ions}, which are cosmochemically abundant, well-known from Earth observations, and readily ionised, but other metal species will also be present.

Section~\ref{ssec:premaven} presents observations of candidate metal ion layers\index{metal layers} throughout the solar system. Section~\ref{ssec:maven} describes {\it MAVEN\index{MAVEN spacecraft}} observations of metal ions at Mars. Section~\ref{ssec:comet} discusses the ionospheric effects of the encounter of Mars with comet C/2013 A1 (Siding Spring)\index{comet C/2013 A1 (Siding Spring)} in 2014.

\subsection{\label{ssec:premaven}Pre-MAVEN Observations of Possible Metal Ion Layers\index{metal layers}}
Radio occultation\index{technique, radio occultation} experiments have been responsible for most measurements of extra-terrestrial ionospheres\index{ionospheres, planetary} \citep[e.g.,][]{withers2010}. These measure vertical profiles of electron density\index{electron density}, but cannot provide any compositional information. They have observed narrow layers of plasma near predicted ablation\index{meteoroid ablation} altitudes on many objects \citep[e.g][]{waite1987, withers2013}.
However, since many processes can produce, transport, and destroy ionospheric plasma, definitive confirmation of the presence of plasma of meteoric origin in an extraterrestrial ionosphere requires the detection of metal ions. Since most ionospheric observing methods detect electrons, rather than ions, direct compositional information is rare for planetary ionospheres\index{ionospheres, planetary}. Furthermore, in situ compositional measurements by retarding potential analyzers and ion mass spectrometers are generally made at altitudes above the putative altitudes of metal ion layers\index{metal layers}.

Here we summarize observations of potential metal ion layers\index{metal layers} in solar system ionospheres\index{ionospheres, planetary}, with the exclusion of observations made at Mars by the {\it MAVEN\index{MAVEN spacecraft}} mission.
Since {\it MAVEN\index{MAVEN spacecraft}} is the only mission to provide relevant ion compositional information, {\it MAVEN\index{MAVEN spacecraft}}'s observations and their implications will be presented in Section~\ref{ssec:maven}.

At Venus, \citet{pesnell2000} suggested that some nightside {\it Pioneer Venus Orbiter\index{Pioneer Venus Orbiter spacecraft}} (PVO) electron density\index{electron density} profiles contain meteoric layers\index{metal layers}, and \citet{butler1976} had earlier suggested that meteoroid influx could produce the surprisingly dense nightside ionosphere. However, recent work has not favored these earlier conclusions \citep{fox1997,withers2008a}.
On the dayside, \citet{witasse2006} suggested that two PVO electron density\index{electron density} profiles near the terminator (Solar Zenith Angle (SZA) of 85.6$^{\circ}$ and 91.6$^{\circ}$) contain meteoric layers \citep{withers2008a}. \citet{withers2013} identified 13 candidate meteoric layers at Venus in {\it Mariner 10}, {\it Venera 9/10}, and {\it Pioneer Venus Orbiter\index{Pioneer Venus Orbiter spacecraft}} data. However, the most convincing detection of a plasma layer whose altitude is consistent with ions derived from meteoroids is that of \citet{paetzold2009}, who used {\it Venus Express\index{Venus Express spacecraft}} radio occultation\index{technique, radio occultation} electron density\index{electron density} profiles. They identified 18 instances of low-altitude plasma layers, but only at solar zenith angles between 55$^{\circ}$ and 90$^{\circ}$. Typical peak plasma densities of $10^{10}$ m$^{-3}$ are reached between 110 and 120 km altitude, peak electron densities increase with decreasing solar zenith angle, and layer shapes are symmetric with respect to peak altitude. This work was based on early {\it Venus Express\index{Venus Express spacecraft}} observations in 2006--2007; a comprehensive survey of the full {\it Venus Express\index{Venus Express spacecraft}} dataset has not yet been published. The in situ {\it Pioneer Venus Orbiter\index{Pioneer Venus Orbiter spacecraft}} Retarding Potential Analyzer (ORPA) and Ion Mass Spectrometer (OIMS) did not report any detections of metal ions at their altitudes above 150 km \citep[e.g.,][and references therein]{brace1983}. 

\begin{figure}[!htbp]
\centering
\figurebox{30pc}{}{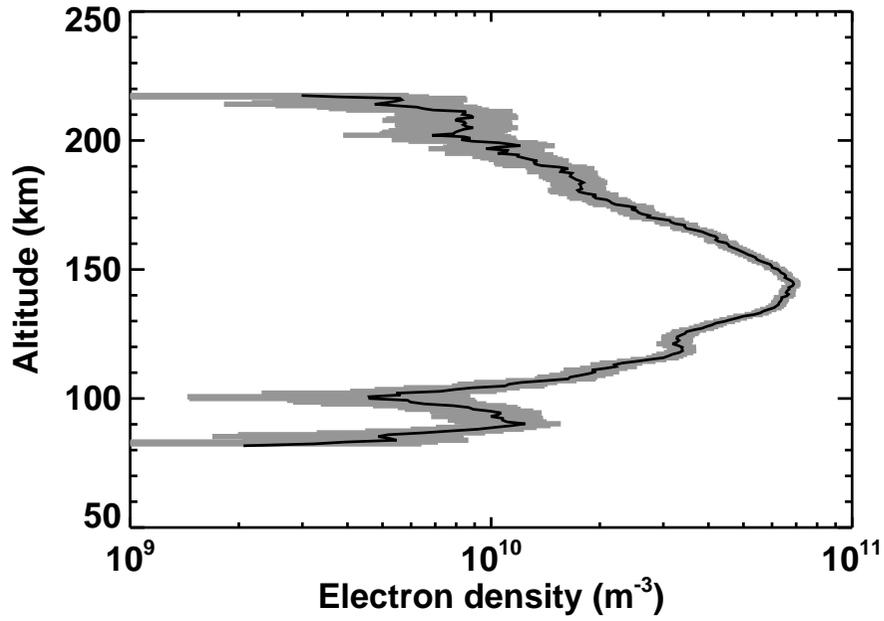}
\caption{Mars Global Surveyor\index{Mars Global Surveyor spacecraft} Radio Science profile 5127R00A.EDS showing a low-altitude plasma layer between 80 and 100 km. It was measured at latitude 66.0${}^{\circ}$N, longitude 2.4${}^{\circ}$E, 14.4 h LST, Solar Longitude = 206.8${}^{\circ}$, and SZA = 81.6${}^{\circ}$ on 2005 May 7. The nominal profile is the solid line, and 1$\sigma$ uncertainties in the electron densities are indicated by the grey region. After \citet{withers2008a}.}
\label{fig:iono}
\end{figure}

At Mars, \citet{fox2004} suggested that a Mars Global Surveyor\index{Mars Global Surveyor spacecraft} electron density\index{electron density} profile displayed a plasma layer near 90 km with a density of $5 \times 10^{9}$ m$^{-3}$ that ``could be attributed to meteoric ions''. \citet{paetzold2005} surveyed 120 {\it Mars Express\index{Mars Express spacecraft}} electron density\index{electron density} profiles and found that 10 of them contained a low-altitude plasma layer between 65 km and 110 km with an average peak electron density\index{electron density} of $8 \times 10^{9}$ m$^{-3}$. Subsequently, \citet{withers2008a} conducted a comprehensive survey of the entire set of 5600 electron density\index{electron density} profiles observed by {\it Mars Global Surveyor\index{Mars Global Surveyor spacecraft}}. They found low-altitude plasma layers in 71 of the 5600 profiles, a significantly lower detection rate than for {\it Mars Express\index{Mars Express spacecraft}}. An example is shown in Figure~\ref{fig:iono}.
The mean altitude of the meteoric layer was 91.7$\pm$4.8 km. The mean peak electron density\index{electron density} in the plasma layer was $\left(1.33 \pm 0.25\right) \times 10^{10}$ m$^{-3}$ and the mean width of the layer was $10.3 \pm 5.2$ km. \citet{pandya2012} investigated these low altitude plasma layers using the integrated total electron content between 80 km and 105 km, found a higher occurrence rate of candidate low-altitude layers than did \citet{withers2008a}, and concluded that the total electron content in this altitude range increased by a factor of 1.5--3.0 when these layers were present. They suggested that some observed increases in total electron content were associated with the predicted occurrence of meteor showers, but did not show that these increases happened in multiple Mars Years. Searches for potential metal ion layers\index{metal layers} in earlier Mars datasets have been inconclusive. Using published images of electron density\index{electron density} profiles, \citet{withers2013} identified one candidate from {\it Mariner 7} and seven candidates from {\it Mariner 9}. Yet when the actual data from the {\it Mariner 9} profiles were acquired and analyzed, \citet{withers2015} concluded that ``no meteoric layers\index{metal layers} have been firmly identified in the {\it Mariner 9} dataset''. The two pre-MAVEN observations of ionospheric composition, made by retarding potential analyzers on the two Viking Landers, did not detect any metal ions in their profiles that extended down to 110 km \citep{hanson1977}.
In a series of conference abstracts, \citet{aikin2005, maguire2006} reported the detection of infrared emissions from the Mg$^{+}$CO$_{2}$ ion, but this work has not yet passed through the peer-review process. We note that considerable effort has been expended to search for correlations between temporal variations in the properties of these low-altitude plasma layers and the predicted occurrence of meteor showers, but no significant relationships have been identified \cite[][and Section~\ref{sec:pred}]{espley2007, withers2008a, pandya2012, withers2013}.

At Jupiter, {\it Pioneer 10\index{Pioneer 10 spacecraft}}, {\it Pioneer 11\index{Pioneer 11 spacecraft}}, {\it Voyager\index{Voyager spacecraft}}, and {\it Galileo\index{Galileo spacecraft}} radio occultation\index{technique, radio occultation} profiles commonly show plasma layers\index{metal layers} that have widths of 10 km and densities of $10^{10}$--$10^{11}$ m$^{-3}$ at altitudes of a few hundred kilometers \citep{fjeldbo1975b,hinson1997,hinson1998,yelle2004}. At Saturn, {\it Pioneer 11\index{Pioneer 11 spacecraft}}, {\it Voyager\index{Voyager spacecraft}}, and {\it Cassini\index{Cassini spacecraft}} radio occultation\index{technique, radio occultation} profiles commonly show narrow plasma layers\index{metal layers} with densities of $10^{10}$ m$^{-3}$ around 1000 km altitude \citep{lindal1985,nagy2006,kliore2009}. Similar narrow plasma layers were also seen by {\it Voyager 2\index{Voyager spacecraft}} at Uranus and Neptune \citep{lindal1987,lindal1992,tyler1989}. {\it Cassini\index{Cassini spacecraft}} was able to make in situ measurements of the composition of Saturn's ionosphere during its proximal orbits, but was not able to search for metal ions. The high speeds of these atmospheric passes restricted the {\it Cassini\index{Cassini spacecraft}} ion mass spectrometer (INMS) to observations at masses below 7 daltons (pers. comm., Moore, 2018).

At Titan, several {\it Cassini\index{Cassini spacecraft}} radio occultation\index{technique, radio occultation} profiles have been classified as ``disturbed'' (T31N, T31X, T57X) \citep{kliore2008,kliore2011}. The two T31 profiles display a broad plasma layer at 500--600 km altitude with peak density of 2--3$\times 10^{9}$ m$^{-3}$. The published T57 profile is not shown below 800 km \citep{kliore2011}. Several hypothesis have been proposed for the plasma layer at 500--600 km, including meteoroid ablation\index{meteoroid ablation} 
\citep{molinacuberos2001,molinacuberos2008} and ion precipitation \citep{cravens2008}. Relative to the neutral scale height, this plasma layer on Titan is somewhat broader than the putative metal ion layers\index{metal layers} identified on other solar system objects. {\it Cassini\index{Cassini spacecraft}} in situ observations by its mass spectrometer (INMS) that extend down to altitudes around 900 km have not detected metal ions 
\citep{vuitton2009,cravens2010}.

No features in the two {\it Voyager 2\index{Voyager spacecraft}} radio occultation\index{technique, radio occultation} electron density\index{electron density} profiles from Triton have been suggested to be metal ion layers\index{metal layers} \citep{tyler1989}. With a surface pressure of 1--2 Pa, it is not clear that Triton's atmosphere is sufficiently dense to ablate meteoroids before they impact the icy surface \citep[e.g.,][]{moses2000}.
No features in radio occultation\index{technique, radio occultation} electron density\index{electron density} profiles of the Galilean satellites have been suggested to be metal ion layers\index{metal layers}.
There are 10 {\it Galileo\index{Galileo spacecraft}} profiles for Ganymede \citep{kliore2001b,kliore2001a}, 6 {\it Galileo\index{Galileo spacecraft}} profiles for Europa \citep{kliore1997}, 8 {\it Galileo\index{Galileo spacecraft}} profiles for Callisto \citep{kliore2002}, and two {\it Pioneer 10\index{Pioneer 10 spacecraft}} and ten {\it Galileo\index{Galileo spacecraft}} profiles from Io \citep{kliore1974, kliore1975, hinson1998b}.
Due to the rarefied neutral atmospheres of these satellites, little ablation\index{meteoroid ablation} of meteoroids will occur prior to meteoroid impact. Furthermore, since these atmospheres are ballistic, not collisional, ablated metal species will not be suspended in the atmosphere (but see Section~\ref{sec:exospheres}).

Narrow plasma layers have been observed below the main ionospheric peak in many solar system ionospheres\index{ionospheres, planetary}. These have often been interpreted as being metal ion layers\index{metal layers} caused by meteoroid ablation\index{meteoroid ablation}, but direct evidence for the presence of metal ions is lacking. It is worth considering why the meteoroid hypothesis for the origin of these plasma layers is quite widely accepted, despite the absence of direct composition measurements.
Two factors seem significant: (1) the existence of meteoroid-derived metal ion layers\index{metal layers} on Earth that appear analogous to these extraterrestrial plasma layers\index{metal layers}; and (2) the paucity of verifiable non-meteoroid explanations for these low-altitude plasma layers. Yet recent observations of ionospheric composition at Mars have shed new light on these issues.

\subsection{\label{ssec:maven}{\it MAVEN\index{MAVEN spacecraft}} Observations of Metal Ions at Mars}
The {\it MAVEN\index{MAVEN spacecraft}} spacecraft is a Mars orbiter that makes extensive measurements of the ionosphere \citep{jakosky2015}. Two {\it MAVEN\index{MAVEN}} instruments are able to detect metal ions. In situ observations by the mass spectrometer \citet[NGIMS;][]{mahaffy2015,benna2015} are sensitive to metal ions above spacecraft periapsis (nominally 150 km, occasionally as low as 120 km) and remote sensing observations by the ultraviolet spectrometer (IUVS, \citep{mcclintock2015}) are sensitive to Mg$^{+}$\index{Mg, ions} ions down to approximately 80 km altitude. Both instruments have detected metal ions.

The NGIMS instrument found that metal ions Na$^{+}$\index{Na, ions}, Mg$^{+}$\index{Mg, ions}, and Fe$^{+}$\index{Fe, ions} are continuously present down to the lowest altitudes sampled (120--130 km) \citep{grebowsky2017}. Densities of Mg$^{+}$\index{Mg, ions} and Fe$^{+}$\index{Fe, ions} are, on average, approximately equal.
This was unexpected -- ``one might expect Fe$^{+}$\index{Fe, ions} to be less dominant with increasing altitude because of the gravitational mass separation anticipated for diffusion processes'' \citep{grebowsky2017}. Ionospheric models also predicted that Mg$^{+}$\index{Mg, ions} would be appreciably more abundant than Fe$^{+}$\index{Fe, ions} \citep{whalley2010}.
However, the Mg$^{+}$/Fe$^{+}$ ratio on an individual orbit may vary significantly from its long-term average. Densities of Mg$^{+}$\index{Mg, ions} and Fe$^{+}$\index{Fe, ions} are, on average, proportional to the density of the dominant neutral constituent, CO$_{2}$. This was also unexpected -- on Earth, ``observations made above the main ionospheric metal ion layer are often characterized by complex layers\index{metal layers} associated with electrodynamic sources, with no clear trend of ordered metal ion concentration decreases with increasing altitude'' \citep{grebowsky2017}. Furthermore, ``isolated metal ion layers\index{metal layers} mimicking Earth's sporadic E layers occur despite the lack of a strong magnetic field as required at Earth'' \citep{grebowsky2017}. These metal ion layers\index{metal layers} are seen at altitudes around 140--170 km and have widths of 10 km.

\begin{figure}[!htbp]
\figurebox{30pc}{}{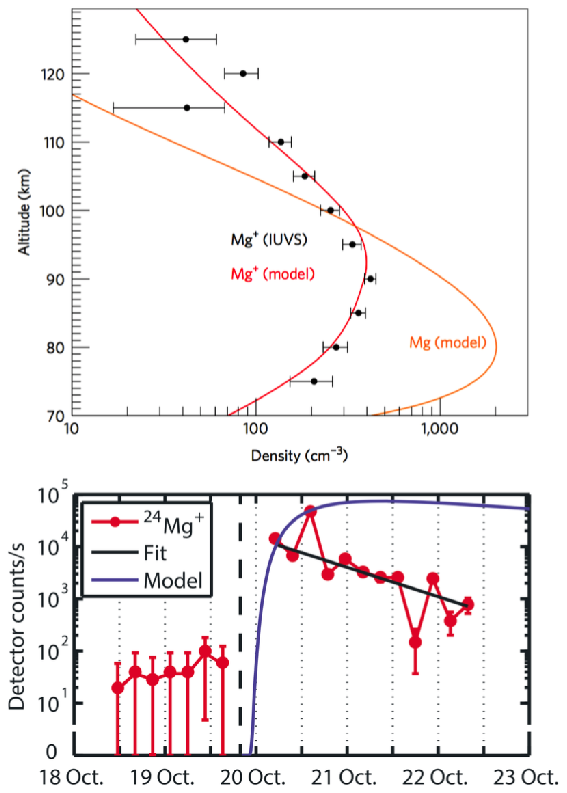}
\caption{{\sl Top}: $\mbox{Mg}^{+}$\index{Mg, ions} altitude profiles derived from {\it MAVEN\index{MAVEN spacecraft}} IUVS at orbit 3040 compared with the model prediction. Note that Mg\index{Mg, neutral} is not detected despite large predicted concentrations.
From \citet{crismani2017}. {\sl Bottom}: Temporal evolution of the abundances of $\mbox{Mg}^{+}$\index{Mg, ions} measured by {\it MAVEN\index{MAVEN spacecraft}}/NGIMS at periapsis from 2014 October 18 to 2014 October 23. The exponential decay can be fitted by a time constant of 1.8 days. The dashed line marks the predicted time of maximum flux of C/2013 A1\index{comet C/2013 A1} dust. Predicted signal levels were derived by a 1-D model. Error bars reflect 3 $\times$ standard deviation of the sampled data due to counting statistics. Instrument background is 10--100 counts/s. From \citet{benna2015}.}
\label{fig:metals}
\end{figure}

The in situ NGIMS observations were unable to address how the trend of exponential increase in metal ion density with decreasing altitude continued below the 120 km limit of periapsis. Remote sensing observations are required to do so. 
The IUVS instrument found that Mg$^{+}$\index{Mg, ions} ions are continuously present in the $\sim$75--125 km altitude range \citep[Fig.~\ref{fig:metals}, top panel;][]{crismani2017}. \citet{crismani2017} reported that Mg$^{+}$\index{Mg, ions} ions formed a layer with peak density $2.5 \times 10^{8}$ m$^{-3}$, peak altitude of 90 km, and full-width at half-maximum of approximately 30 km. The layer shape appears symmetric with distance from the peak, distinct from the strikingly asymmetric shape of a Chapman layer.

Based on \citet{grebowsky2017} and \citet{crismani2017}, the NGIMS and IUVS observations of Mg$^{+}$\index{Mg, ions} appear to be consistent, but further work is needed to synthesize them into a coherent picture of the vertical profile of Mg$^{+}$\index{Mg, ions} density from 75 km to 175 km.

Both radio occultation\index{technique, radio occultation} experiments and {\it MAVEN\index{MAVEN spacecraft}} IUVS have observed a plasma layer at 90 km altitude. However, radio occultation\index{technique, radio occultation} experiments observe a sporadic layer of electrons with peak density in excess of $10^{10}$ m$^{-3}$ and width of 10 km, whereas IUVS observes a continuous layer of Mg$^{+}$\index{Mg, ions} ions with peak density that is 50 times smaller and width that is three times larger. 
The persistent presence of Mg$^{+}$\index{Mg, ions} ions in IUVS observations is consistent with terrestrial experience, where metal ions are a ubiquitous feature of the ionosphere. They derive predominantly from steady influxes of sporadic meteoroids, rather than shower meteoroids whose influx varies greatly with time. Shower meteoroids, although visually striking, contribute only a small fraction of the steady-state mass flux \cite[e.g.][]{Brown2008}. Readers should note the potential for confusion between the ``sporadic'' occurrence of low altitude electron density\index{electron density} layers in radio occultation\index{technique, radio occultation} profiles and the ``sporadic'' meteoroid population, which supplies a relatively stable influx of interplanetary dust.

If the sporadic low-altitude plasma layers seen by radio occultation\index{technique, radio occultation} experiments are generated by meteoroid influx, then their chemistry must be such that metal ions like Mg$^{+}$\index{Mg, ions} rapidly transform into other ion species that are much longer-lived. That possibility is extremely inconsistent with current understanding of the effects of meteoroids on planetary atmospheres \citep[e.g.,][]{whalley2010}. 
Furthermore, their irregular occurrence must be reconciled with the constant influx of interplanetary dust. The other possibility is that the sporadic low-altitude plasma layers seen by radio occultation\index{technique, radio occultation} experiments are caused by other mechanisms, such as enhancements in the precipitation rate of charged particles of suitable energy to deposit their energy at these altitudes.

In light of the {\it MAVEN\index{MAVEN spacecraft}} IUVS observations, we judge that the cause of the sporadic low-altitude plasma layers seen at Mars by radio occultation\index{technique, radio occultation} experiments is currently unknown. Given the similarities between the ionospheres\index{ionospheres, planetary} of Venus and Mars, caution should also be applied to the interpretation of the analogous layers in the ionosphere of Venus. However, given the many major differences between giant planets and terrestrial planets, these Mars findings should not necessarily be extended directly to the giant planets. Nevertheless, it is important to note that direct evidence for metal ions in giant planet ionospheres\index{ionospheres, planetary} is absent.

\subsection{\label{ssec:comet}Encounter of Mars with Comet C/2013 A1 (Siding Spring)\index{comet C/2013 A1 (Siding Spring)}}
On 2014 October 19, Mars experienced a remarkably close encounter with comet C/2013 A1 (Siding Spring)\index{comet C/2013 A1 (Siding Spring)} at a distance of approximately 135,000 km \citep[e.g.,][and references therein]{withers2014}. As Mars passed through the comet's coma, the dust influx on Mars increased substantially. This was expected to affect the distribution of metal ions in the planet's ionosphere. This encounter occurred when several spacecraft were operational at Mars and able to perform relevant observations. {\it MAVEN\index{MAVEN spacecraft}}, which arrived at Mars only weeks earlier, searched for metal ions using the NGIMS and IUVS instruments discussed above. The MARSIS topside radar sounder on {\it Mars Express\index{Mars Express spacecraft}} measured the peak density of the ionosphere. The SHARAD radar sounder on Mars Reconnaissance Orbiter\index{Mars Reconnaissance Orbiter spacecraft} measured the total electron content of the ionosphere.

{\it MAVEN\index{MAVEN spacecraft}} IUVS observed enhanced densities of Mg$^{+}$\index{Mg, ions} (Figure~\ref{fig:metals}, bottom panel) and Fe$^{+}$\index{Fe, ions} ions after the encounter \citep{schneider2015}. (Note that Fe$^{+}$\index{Fe, ions} was not discussed by \citet{crismani2017} in their survey of the long-term behavior of Mg$^{+}$\index{Mg, ions} ions.)
A typical vertical profile of Mg$^{+}$\index{Mg, ions} density several hours after closest approach revealed a layer with a peak density on the order of 
$10^{10}$ m$^{-3}$, peak altitude of 120 km, symmetric shape, and full-width at half-maximum of approximately 10 km. Relative to observations under normal conditions, the peak density is larger, the peak altitude is higher, and the width is narrower. The increase in peak density is expected from the increase in dust flux. The increase in peak altitude is expected from the increase in dust speed from sporadic meteoroids at a few km\,s$^{-1}$ to cometary dust at 56 km\,s$^{-1}$. The decrease in peak width has not been explained. \citet{schneider2015} found that Mg$^{+}$\index{Mg, ions} densities were enhanced globally for 1--2 days.

The {\it MAVEN\index{MAVEN spacecraft}} NGIMS instrument observed enhanced densities of a tremendous range of metal ions (singly ionised Na\index{Na, ions}, Mg\index{Mg, ions}, Al, K, Ti, Cr, Mn, Fe\index{Fe, ions}, Co, Ni, Cu, and Zn, and possibly Si and Ca\index{Ca, ions}) at 185 km after the encounter \citep{benna2015}. Observations did not extend below 185 km, which was the periapsis altitude at this early stage of the mission. In these observations, the most abundant metal ions were Na$^{+}$\index{Na, ions}, Mg$^{+}$\index{Mg, ions}, and Fe$^{+}$\index{Fe, ions}, but in different proportions to those seen in normal circumstances. Here Na$^{+}$\index{Na, ions} was the most abundant ion (roughly 3 times as abundant as Fe$^{+}$\index{Fe, ions}) and Mg$^{+}$\index{Mg, ions} was the next most abundant (roughly 2 times as abundant as Fe$^{+}$\index{Fe, ions}). That may reflect compositional differences between fresh dust from comet C/2013 A1 (Siding Spring)\index{comet C/2013 A1 (Siding Spring)} and background interplanetary dust. The Mg$^{+}$\index{Mg, ions} density at 185 km shortly after the encounter ($153.6\pm7.5 \times 10^{6}$ m$^{-3}$) was increased from its normal value by approximately three orders of magnitude, which illustrates the dramatic impact of this comet on the distribution of metal species in the ionosphere and atmosphere\index{atmosphere, Mars}. The Mg$^{+}$\index{Mg, ions} density at 185 km decreased exponentially after the encounter with a time constant of 1.8 days. 

Electron densities in the ionosphere of Mars were observed by other orbital instruments during the encounter. The MARSIS topside radar sounder on {\it Mars Express\index{Mars Express spacecraft}} observed peak electron densities of 1.5--2.5$\times 10^{11}$ m$^{-3}$ at 80--100 km altitude and solar zenith angles of 75--110$^{\circ}$ \citep{gurnett2015}. Under normal conditions, the peak electron density\index{electron density} and peak altitude at these solar zenith angles are 0.1--1.0$\times 10^{11}$ m$^{-3}$ and greater than 140 km, respectively. Although the ionospheric composition was not simultaneously observed for these electron density\index{electron density} observations, it seems likely that metal ions constituted a significant fraction of the total ion density.

The SHARAD radar on Mars Reconnaissance Orbiter\index{Mars Reconnaissance Orbiter spacecraft} measured the vertical total electron content of the ionosphere at solar zenith angles of 94$^{\circ}$ and 113$^{\circ}$ shortly after the encounter \citep{restano2015}. It found total electron content values on the order of 4$\times 10^{15}$ m$^{-2}$ (94$^{\circ}$) and 2$\times 10^{15}$ m$^{-2}$ (113$^{\circ}$). Under normal conditions, total electron contents at these solar zenith angles are approximately 2$\times 10^{15}$ m$^{-2}$ (94$^{\circ}$) and 0.5$\times 10^{15}$ m$^{-2}$ (113$^{\circ}$). The corresponding peak electron density\index{electron density} can be estimated from the fact that the total electron content is usually on the order of $4NH$ where $N$ is the peak electron density\index{electron density} and $H$ is the neutral scale height (10 km) \citep{withers2011}. The inferred peak electron densities are $10^{11}$ m$^{-3}$ (94$^{\circ}$) and $5 \times 10^{10}$ m$^{-3}$ (113$^{\circ}$), which are reasonably consistent with the MARSIS peak density measurements when allowance is made for plausible spatial and temporal variations in ionospheric conditions during this unusual event.

The overall picture of the ionosphere of Mars suggested by the {\it MAVEN\index{MAVEN spacecraft}} IUVS, {\it MAVEN\index{MAVEN spacecraft}} NGIMS, {\it Mars Express\index{Mars Express spacecraft}} MARSIS, and Mars Reconnaissance Orbiter\index{Mars Reconnaissance Orbiter spacecraft} SHARAD observations is of enhanced plasma densities at relatively low altitudes, altered ion composition, and substantial spatial and temporal variations. Realistic numerical simulations of the global-scale, time-varying picture of ionospheric properties during the cometary encounter have not yet been conducted. Such simulations, though technically challenging, are necessary to synthesize the disparate measurements available for this unique event.

In conclusion, metal ion layers\index{metal layers} are likely to exist in all planetary ionospheres\index{ionospheres, planetary}, but the narrow electron density\index{electron density} layers that have been observed throughout the solar system and suggested as being meteoric have not been proven to contain metal ions. In view of the difficulty of in situ sampling of composition at relevant pressure levels, remote sensing by ultraviolet spectroscopy offers the most promising path for detecting metal ions in planetary ionospheres.

\section{\label{sec:flashes}Impact Flashes\index{impact flashes, Jupiter} at Jupiter}
The giant planet Jupiter is observed every year by thousands of amateur astronomers who acquire 
video\index{video technique} observations of its atmosphere\index{atmosphere, Jupiter}. Their images provide a nearly continuous observational record 
that is widely used to study the atmospheric dynamics of the planet \citep{Hueso10a,Hueso18a,Mousis14}. 
Since the year 2010 some of these video\index{video technique} observations have resulted in the serendipitous discovery of energetic flashes\index{impact flashes, Jupiter} of light that last from 1 to a few seconds and with visual brightness comparable to stars of magnitude +6$^{\rm m}$ or brighter. 
Although Jupiter impacts visible from Earth are caused by asteroids or comets, not meteoroids, they do nevertheless produce meteor phenomena. The first of these optical flashes\index{impact flashes, Jupiter} was discovered on 2010 June 3 and was simultaneously recorded by two observers (Anthony Wesley from Australia discovered the flash and Christopher Go from Philippines confirmed this with his own observation acquired at the same time). This bolide appeared as a bright flash\index{impact flashes, Jupiter} that lasted about 2 seconds and did not leave any observable trace on the atmosphere\index{atmosphere, Jupiter} afterwards. A large follow-up campaign of observations using telescopes such as the Very Large Telescope (VLT) or Hubble Space Telescope (HST) did not find any debris in the atmosphere \citep{Hueso10b}. The analysis of the light-curve of the observations yielded an estimated energy of the impact of (1.9--14) $\times 10^{14}$ J which in turn corresponds to an object of 1.05--7.80 $\times 10^{5}$ kg with a diameter range of 4.7--18 m depending on the density \citep{Hueso10b, Hueso13}.

Since then, other instances of energetic flashes\index{impact flashes, Jupiter} on Jupiter have been found, totalling 5 different events observed by a total of 12 amateur observers. The light-curves of these events have been analyzed by \cite{Hueso13} and \cite{Hueso18b} and interpreted as being caused by the impact of large meteoroids with diameters of 5--20 m. We note that these actually qualify as small asteroids or cometary fragments rather than meteoroids according to the IAU definition\footnote{https://www.iau.org/public/themes/meteors\_and\_meteorites/}. The estimated masses and energies released in Jupiter's atmosphere by these impacts are summarized in Table~\ref{tab:Impacts} where we also compare with previous impacts in Jupiter: the meteor observed by {\it Voyager 1\index{Voyager spacecraft}} in 1981 \citep{Cook81}, the impacts of the D/1993 F2 (Shoemaker-Levy 9) fragments in 1994 \citep{Hammel95, Harrington04} and the 500-m diameter object that impacted Jupiter in 2009 \citep{SanchezLavega10}, the latter also recorded by Anthony Wesley one year before his discovery of the bright flash.

Remarkably, each of these bolides was first noted by a single observer who detected the flash\index{impact flashes, Jupiter} and issued an alert. The alert quickly resulted in reports from other observers who found the flashes\index{impact flashes, Jupiter} in their video\index{video technique} observations acquired on the same night and at the precise time of the reported flash\index{impact flashes, Jupiter}. This occurs because most video\index{video technique} recordings are obtained over several minutes and it is difficult to manually scrutinise video\index{video technique} observations frame by frame in search of an unremarkable short-lived feature such as a flash\index{impact flashes, Jupiter}. Software tools that process video\index{video technique} observations to detect flashes\index{impact flashes, Jupiter} are now available at, eg, 
\url{http://www.astrosurf.com/planetessaf/doc/project_detect.php} and \url{http://pvol2.ehu.eus/psws/jovian_impacts/} \citep{Andre18}. However, the availability of these tools has not produced further detections, probably due to their limited usage up to now.

\begin{figure}[!htbp]
\centering
\figurebox{30pc}{}{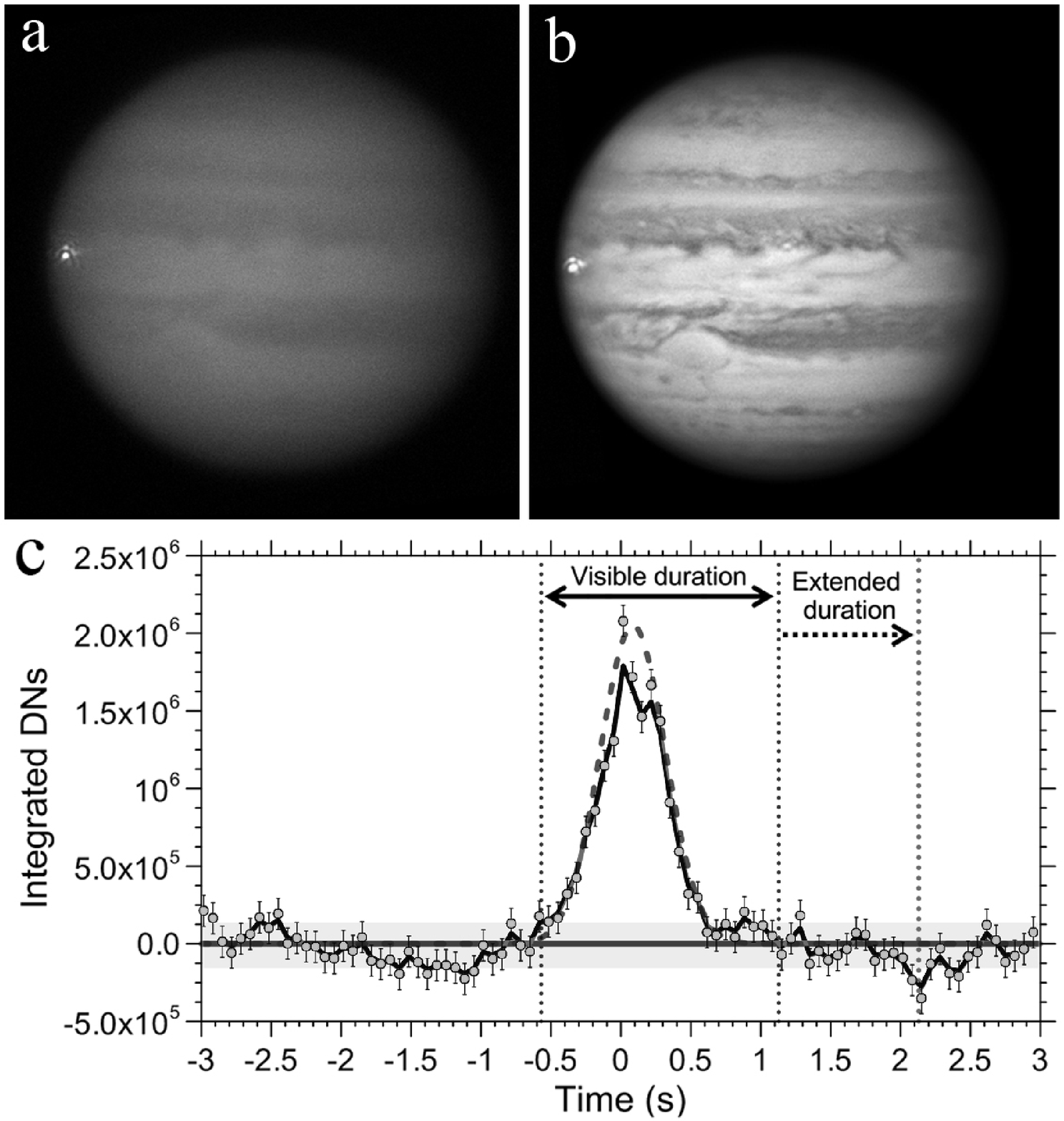}
\caption{The 2012 September impact on Jupiter. This was the most energetic flash\index{impact flashes, Jupiter} of light at Jupiter. (a) Brightest video\index{video technique} frame in the recording; (b) Image composite made by stacking all frames where the flash\index{impact flashes, Jupiter} is visible and 1000 frames for the planet. (c) Light-curve of the impact. The impact partially saturated some pixels at the peak of its brightness.
Points represent the flash\index{impact flashes, Jupiter} intensity measured over the different frames, the dashed line is a gaussian fit to the data and the continuous line is a weighted average of the photometric values. Further evidence of the flash\index{impact flashes, Jupiter} is visible in the data at the level of the photometric noise. Based on data from \citet{Hueso13}.}
\label{fig:Impact2012}
\end{figure}

The brightest flash\index{impact flashes, Jupiter} occurred in 2012 September and is discussed in \cite{Hueso13}. Figure~\ref{fig:Impact2012} shows the frame with the peak of the flash\index{impact flashes, Jupiter}, a synthetic image of the \index{impact flashes, Jupiter} made by stacking the different frames where the flash\index{impact flashes, Jupiter} is visible and a light-curve of this event. Analysis of the light-curve following photometric calibration of the video\index{video technique} indicates a mass of about 5.00--9.50 $\times 10^{5}$ kg for the impactor. 

\cite{Hueso13} present a numerical simulation of the airbursts produced by objects equivalent to the 2012 impact and colliding with Jupiter at velocities of 60 km\,s$^{-1}$. Objects of this size begin to break up at about 3 mbar pressure (120 km above the 1-bar level) and deposit most of their energy at the 5--6 mbar level (100 km height) where they break up in successive events. These result in light-curves that may show significant structure if observed with sufficient time resolution.

The latest analysis of the frequency of such impacts on Jupiter results in an estimate of 4--25 detectable events per year \citep{Hueso18b}. Given that these flashes\index{impact flashes, Jupiter} have to be found right at the moment they occur, and we may only observe one side of Jupiter and for a limited period of time every year,  the total number of events on Jupiter's atmosphere should be 10--70 impacts per year. This impact rate is in basic agreement with expectations from dynamical studies of comets \citep{Levison00}. The contribution of impacts of this size range to the abundance of well-characterized exogenic species (mainly O atoms in CO and H$_{2}$O molecules; \citealt{Lellouch02,Moses17}) in Jupiter's upper atmosphere\index{atmosphere, Jupiter} is probably limited \citep{Hueso18a}.

We expect that the number of detections in the next few years will increase. On the one hand, the new generation of sensitive cameras used by amateur astronomers will allow to detect impacts by slightly smaller objects. On the other hand, Jupiter oppositions in the next few years will occur in spring and summer in the North hemisphere, where most of the regular observers of Jupiter are located. Additionally, improvements in the software tools used by the amateur community to systematically search for more flashes\index{impact flashes, Jupiter} are underway, and will allow to improve the detectability of these events, providing better constraints on the flux of impacts in Jupiter. Finally, similar impacts are also expected to occur on Saturn \citep{Tiscareno13}. Their discovery from ground-based observations is not beyond the capability of equipment used by amateurs to discover bolide impacts on Jupiter.
\section{\label{sec:exospheres} Impact-Derived Species Within Airless Body Exospheres\index{surface boundary exosphere}}
Objects such as the Moon, Mercury and asteroids that lack atmospheres in the traditional sense, host instead tenuous Surface-Bounded Exospheres\index{surface boundary exosphere} (SBEs), the result of a delicate balance between poorly-understood sources and sinks.

Models of SBEs have extremely varied estimates of the importance of surface Impact Vapourisation\index{impact vapourisation} (IV). Although \index{impact vapourisation} is an established field of study \citep{Melosh1989,Pierazzo2008,Hermalyn2010} uncertainties regarding the importance of impact vapourisation\index{impact vapourisation} on extraterrestrial bodies include the uncertainty in impact rates for both interplanetary dust and larger meteoroids and comets \citep{Cremonese2013}, the relative amounts of melt and vapor produced in an impact \citep{Pierazzo1995}, the temperature of the vapor -- which affects escape rates \cite[e.g. ][]{Cintala1992,Rivkin2005}, the relative amount of neutral vs. ionised ejecta \citep{Hornung2000}, and the gas-surface interaction of the downwelling ejecta \citep{Yakshinskiy2005}. The importance of impact vapourisation\index{impact vapourisation} as a source of exospheric neutrals has been constrained in part by observation of the escaping component of the exospheres\index{surface boundary exosphere} -- the Mercurian tail \citep{Schmidt2012}, the Ca\index{Ca, neutral} exosphere\index{surface boundary exosphere} of Mercury \citep{Burger2014} and the lunar extended exosphere\index{surface boundary exosphere} and tail \citep{Wilson1999}. These results all depend critically on the assumed temperature or velocity distribution of the initial vapor plume, which has been variously assumed to be between 1500 K to 5000 K or non-thermal. They also depend critically on the assumed photoionisation rate. Values of the Na\index{Na, neutral} photoionisation rate have varied by a factor of three, and values of the Ca photoionisation rate have varied by a factor of 4.4 (e.g. see \cite[e.g. see ][]{Killen2019}. This obviously introduces a huge uncertainty in the escape rate, and hence the source process.

\citet{Colaprete2016} conclude, based on observations of the UV spectrometer onboard the {\it LADEE\index{LADEE spacecraft}} mission, that there is a pronounced role for meteoroid impact vapourisation\index{impact vapourisation} and surface exchange in determining the composition of surface-bounded exospheres\index{surface-bounded exospheres}. Most interestingly, the simulations show that a release of ejecta from a single injection will persist in the SBE-surface system for much longer than the ionisation lifetime. Their nominal model shows that about 30\% of Na\index{Na, neutral} released is still adsorbed on the surface after 100 days (3 lunations). Residence times in the lunar environment of 45 to 90 days (mainly on the lunar surface) can be expected before escape to the solar wind, which would explain the long-term smooth increase and decrease in the Na\index{Na, neutral} column density observed as the result of meteoroid streams\index{meteoroid streams}. This is the result of each particle residing in the soil for approximately an ionisation lifetime (i.e., several days) between bounces, combined with the many bounces that it has to take before being lost from the SBE. Although \citet{Leblanc2010} conclude that thermal desorption and Photon-Stimulated Desorption (PSD) are the dominant source processes for the Na\index{Na, neutral} exosphere\index{surface boundary exosphere} of Mercury, they introduce a source term equivalent to the amount of fresh material required to maintain the SBE. \adjustfigure{100pt} This source is not identified but is required to maintain a surficial reservoir of adsorbed atoms containing at least 2000 times as many atoms as the exospheric content.

A pronounced increase in K was observed by {\it LADEE\index{LADEE spacecraft}} during the Geminid meteor shower while no increase was seen in Na\index{Na, neutral}. Given that Na\index{Na, neutral} is more volatile than K, it is not obvious that the peaks are the result of the meteoroid streams\index{meteoroid streams} as opposed to whatever is causing the monthly variation. Given the long residence time of Na\index{Na, neutral} on the surface deduced by both \citet{Leblanc2003} and \citet{Colaprete2016}, it has been suggested that impacts are the primary source of atoms from the regolith to the extreme surface, and these atoms feed the subsequent release by photons or thermal processes. If this is the case, then micrometeoritic impacts play the dominant role in maintaining the SBEs while the less energetic processes such as PSD and thermal desorption serve to keep the atoms in play until they are destroyed by photoionisation.

\citet{Schmidt2012} studied the extended sodium tail of Mercury and concluded that both photon-stimulated desorption and micrometeoroid impacts are required to simulate the $\sim$20\% loss of Mercury’s sodium atmosphere, depending on orbital phase, and that the two mechanisms are jointly responsible for the observed comet-like tail as driven by solar radiation pressure. Roughly three times as many atoms would have to be ejected by PSD than by IV at 3000 K to accomplish a similar loss rate. The velocity distribution for PSD was required to have a low energy maximum (900 K) and high energy tail similar to that measured by \citet{Yakshinskiy2004} for electron-stimulated desorption from a lunar sample at 100 K.

The distant sodium tail of the Moon has been observed by looking in the direction opposite the sun \citep{Wilson1999}. They concluded that the Na\index{Na, neutral} escape rate increased by a factor of 2 -- 3 during the most intense period of the 1998 Leonid meteor shower. The changes in the lunar exosphere\index{surface boundary exosphere} itself were not quantified, as the escape rates only constrain that fraction of the velocity distributions above 2.1 km\,s$^{-1}$. Nevertheless, the observation is evidence for a strong influence of meteor impacts on the lunar sodium exosphere\index{surface boundary exosphere} and its escape rate.

\begin{figure}[!htbp]
\figurebox{30pc}{}{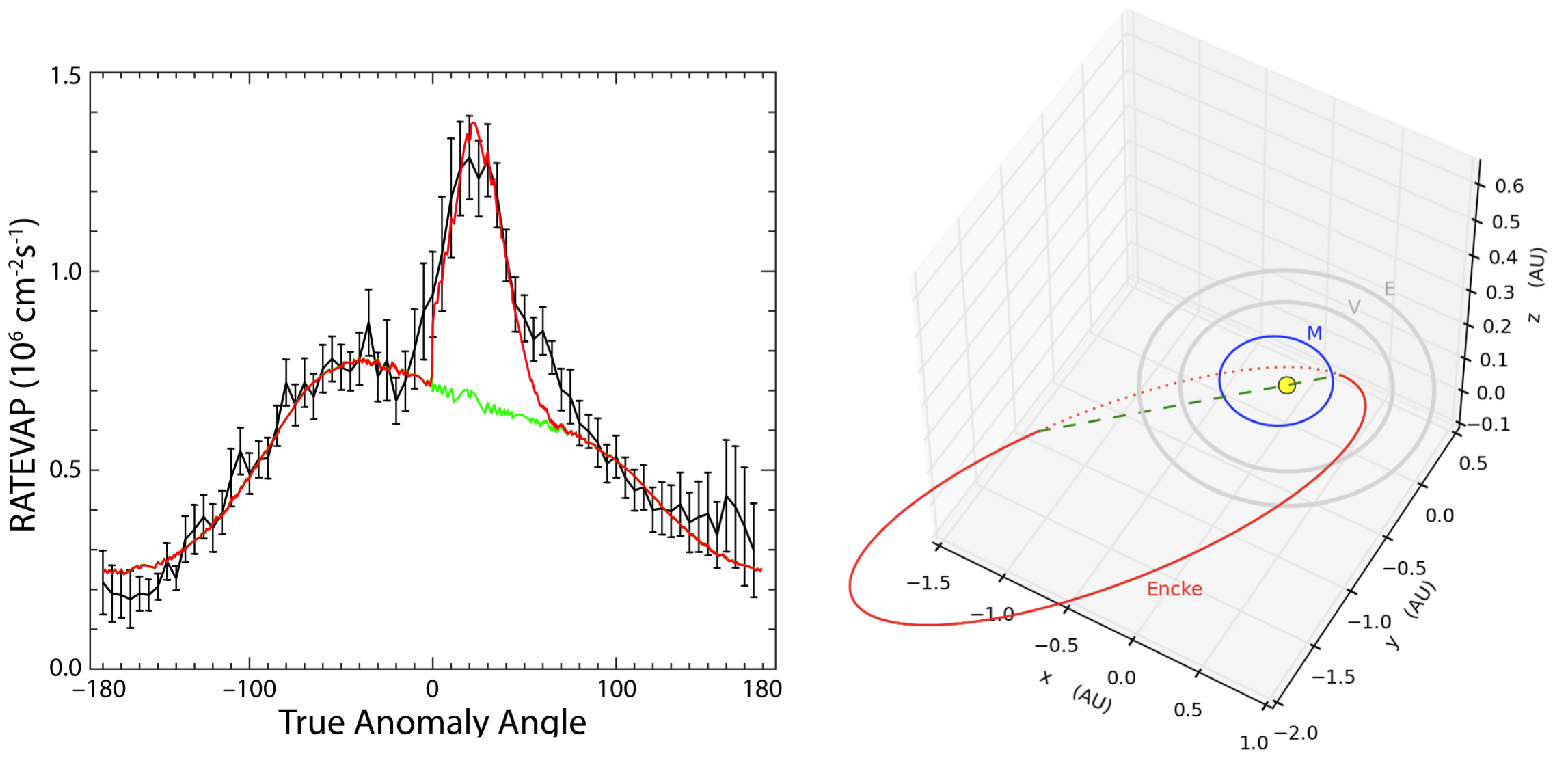}
\caption{{\sl Left panel}: Total planetary calcium source rate to Mercury's exosphere\index{surface boundary exosphere} is a periodic function of the planet's TAA. Mercury is at perihelion when TAA=0$^{\circ}$ and at aphelion when TAA=$\pm$ 180$^{\circ}$. The black curve is this total rate summed over the planet at each TAA, derived from observations obtained by the {\it MESSENGER\index{MESSENGER spacecraft}} MASCS spectrometer 2011 March -- 2013 March \cite[from][]{Burger2014}. The red line is the modeled contribution from a cometary dust stream with peak density at TAA=25$^{\circ}$ plus that due to an interplanetary dust-disk. The green line is the contribution from the disk. Adapted from \citet{Killen2015}. {\sl Right panel}: Encke's orbit (red) along with those of Mercury (blue), Venus and Earth (in grey). After \citet{Killen2015}.}
\label{fig:exosphere}
\end{figure}

The calcium exosphere\index{surface boundary exosphere} of Mercury exhibits several attributes that point to meteoritic impact as the source of the calcium exosphere\index{surface boundary exosphere} (Figure~\ref{fig:exosphere}). First, the calcium source peaks strongly on the dawn side of the planet, exhibiting the same morphology seen in the lunar dust observed by the {\it LADEE\index{LADEE spacecraft}} spacecraft \citep{Szalay2018,Szalay2015,Janches2018}. The variation of the calcium exosphere\index{surface boundary exosphere} with True Anomaly Angle (TAA) was modeled independently by \citet{Killen2015} and \citet{Pokorny2018}, indicating that the calcium source rate could be the result of impacts from the interplanetary dust disk except for a dramatic increase near TAA of 20$^{\circ}$ -- 30$^{\circ}$. This increase was attributed to the intersection of Mercury's orbit and that of comet Encke \citep{Killen2015} and further shown to be consistent with the evolution of the dust stream by the influence of planetary perturbations and Poynting-Robertson drag \citep{Christou2015}.

%\adjustfigure{120pt}
Although much of the literature concerning impact vapourisation\index{impact vapourisation} on airless bodies relates to the ejection of sodium, potassium or calcium, the origin of schreibersite, a phosphide common to impact breccias at all Apollo sites, has been proposed to be a meteoritic contaminant, or alternatively produced {\sl in situ} by reduction on the lunar surface. \citet{Pasek2015} proposed that schreibersite and other siderophilic P phases have an origin from impact vapourisation\index{impact vapourisation} of phosphates at the lunar oxygen fugacity. Phosphorus has not been observed in any SBE but that may be due to the difficulty in observing it.
\section{\label{sec:future}Exo-Meteor Observations: Future Prospects}
In the near future, we will continue to rely on serendipity for {\sl in situ} detection of exo-meteors, either directly or indirectly. The {\it MAVEN\index{MAVEN}} orbiter continues to monitor the metal content of Mars' upper atmosphere for increases of similar magnitude and character to the C/2013 A1 (Siding Spring) event that would indicate passage through a comet's dust stream. At the surface, the {\it Opportunity\index{Mars Exploration Rovers}} and {\it Curiosity\index{Curiosity rover}} rovers regularly image the Martian sky but confirmation of a martian meteor will require the fortuitous interruption of the trail by a foreground object, such as a rock, a mountain or part of the spacecraft structure \citep{Christou.et.al2007b}. The {\it InSight\index{InSight lander}} lander scheduled to land on the Martian surface in late 2018 will be equipped with seismometers that will detect atmospheric entry of large meteoroids \citep{Stevanovic.et.al2017}. Optical cameras are also present on the spacecraft, but are not designed for night time observations and therefore will most likely not deliver many meteor images.
Closer to the Sun, the JAXA {\it Akatsuki\index{Akatsuki spacecraft}} spacecraft finally entered orbit around Venus in 2015 November, at the end of an extended cruise period. The Lightning and Airglow Camera (LAC) is a $12^{\circ}$ FOV high-speed imaging system onboard the orbiter that searches for rapidly-varying luminous phenomena in the night-time Venusian atmosphere \citep{Takahashi.et.al2008}. The instrument has been active since 2016 December, observing Venus for $\sim$30min every 10-day orbit \citep{Takahashi.et.al2017} and may detect bright meteors if, or when, their frequency is sufficiently high. In the meantime, valuable operational experience in this type of observation is being gained by ongoing -- as well as planned -- spaceborne monitoring of the atmosphere of our own planet for meteor events \citep{Arai2014,Abdellaoui.et.al2017}.
These efforts are complemented by ground-based monitoring of the atmosphere of Jupiter by amateur and professional astronomers for impact flashes\index{impact flashes, Jupiter}, possibly to be extended to the other giant planets as well as the Venusian nightside atmosphere.
Numerical modeling and supporting laboratory work is also likely to continue. Many aspects of recent {\it MAVEN\index{MAVEN spacecraft}} observations of metal ions at Mars have yet to be reproduced or explained by numerical models. In particular, rate coefficients for many relevant chemical reactions involving metal species and ambient atmospheric molecules are not well-known. Previous laboratory work has naturally focused on reactions involving the nitrogen and oxygen species prevalent in Earth's atmosphere, not the carbon and hydrogen species common in other atmospheres.
%\appendix
%\section{Copyright Information}
% \backmatter
%% Using an acknowledgements command is not in the Elsevier template,
%% but it can be used.

\contributor{apostolos christou}
{Armagh Observatory and Planetarium\\
				College Hill, BT61 9DG\\ 
Northern Ireland, UK}
\contributor{jeremie vaubaillon}
{IMCCE, Observatoire de Paris\\ 
                          Paris 75014, France}
\contributor{paul withers}
{Astronomy Department, Boston University\\
                   725 Commonwealth Avenue\\ 
                    Boston MA 02215, USA}
\contributor{ricardo hueso}
{Fisica Aplicada I \\
 Escuela de Ingenieria de Bilbao\\ 
 Plaza Ingeniero Torres Quevedo\\ 
 1 48013 Bilbao, Spain} 
\contributor{rosemary killen}
{NASA/Goddard Space Flight Center\\
Planetary Magnetospheres, Code 695 \\
                  Greenbelt MD 20771, USA}           
% \label{lastpage}

\bibliography{chapter5}
%% Use the plainnat style for ``Icarus'' mode to display DOI numbers
%% among other things.  However, revert to the Elsevier elsart-harv
%% mode for ``Elsevier'' mode.
\bibliographystyle{cambridgeauthordate-six-au}
% \bibliographystyle{elsart-harv}
%% --Tables-- 
% \clearpage	% Make sure things don't run together.
% \protect
% \listoffigures
\clearpage
% Table of forecasted meteor showers at Mars
%% --Figures-- %%
% ----------------
% ----------------
% ----------------

\begin{table}
\caption{\label{tab:futureMSMars}Meteor showers at Mars in the next few years, from the method described in \citet{Vaubaillon.et.al2005a} and \citet{VaubaillonChristou2006}. ``parent body'' is the parent body of the stream causing the shower, ``d'' the minimum distance\index{orbit-to-orbit distance} between the center of the stream cross-section and the planet (negative if the stream crosses inside the planet's orbit and positive otherwise), ``Date'' the epoch of the shower, $\alpha$ and $\delta$ the sky location of the radiant in J2000 Earth equatorial coordinates, $V_{p}$ the planetocentric velocity and ``conf\_id'' the confidence index as defined by \citet{Vaubaillon2017}.}
\begin{tabular}{@{}l@{ }rccrccr@{}}
\toprule
parent       & d       & Date        & $\alpha$  & $\delta$&$V_{p}$&      &         \\
body       & (au)       & (UT)        & (deg)  & (deg) &(km\,sec$^{-1}$)&  ZHR     &   conf\_id       \\
\hline
252P/  & $-$0.0260 & 2019-11-17T07:32 &  70.2 &   $-$45.0 &   14.8 &    28   & GYO1/60CU22.8 \\
3D/     &  0.0004 & 2019-12-11T22:27 &  20.1 &    $-$2.0 &   22.6 &    22   &   SYO0/1CE0.0 \\
49P/  & $-$0.0006 & 2019-06-11T13:02 &  22.3 &   $-$67.0 &   16.4 &  1164   &   SYO1/1CE0.0 \\
P/2012 S2 &  0.0616 & 2020-12-23T07:44 & 254.8 &   $-$40.8 &   16.4 &   917   &   SYO0/1CE0.0 \\
156P/      & $-$0.0856 & 2020-11-06T13:56 & 311.5 &   $-$79.5 &   12.6 &    67   &  GYO0/43CU0.0 \\
156P/      & $-$0.0807 & 2020-11-06T20:36 & 312.4 &   $-$79.7 &   12.6 &    25   &   SYO0/1CE0.0 \\
156P/      & $-$0.0855 & 2020-11-06T20:31 & 314.8 &   $-$79.7 &   12.6 &   115   &   SYO0/1CE0.0 \\
156P/      & $-$0.0970 & 2020-10-31T14:47 & 292.3 &   $-$74.1 &   11.6 &   138   &   SYO0/1CE0.0 \\
156P/      & $-$0.0698 & 2020-11-06T09:52 & 320.1 &   $-$81.8 &   13.1 &    57   &   SYO0/1CE0.0 \\
10P/      & $-$0.0358 & 2020-06-16T05:31 & 244.8 &    25.6 &   10.4 &   231   & GYO20/56CU0.0 \\
C/2007 H2\index{comet C/2007 H2 (Skiff)}  &  0.0009 & 2021-10-28T16:58 &  39.9 &   $-$55.2 &   34.4 &   122   &   GYO0/4CU0.0 \\
304P/         &  0.0035 & 2021-01-25T01:47 & 280.8 &   $-$19.1 &   13.5 &   610   &   SYO0/1CE0.0 \\
304P/         & $-$0.0057 & 2021-01-31T01:44 & 280.8 &   $-$19.3 &   13.9 &  1640   &   SYO0/1CE0.0 \\\botrule
\end{tabular}
\end{table}
\clearpage

\begin{table}
\caption{\label{tab:futureMSVenus}Future meteor showers at Venus. See Table~\ref{tab:futureMSMars} for explanations.}
\begin{tabular}{@{}l@{ }rccrccr@{}}
\toprule%
parent       & d       & Date        & $\alpha$  & $\delta$&$V_{p}$&      &         \\
body       & (au)       & (UT)        & (deg)  & (deg) &(km\,sec$^{-1}$)&  ZHR     &   conf\_id       \\
\hline
2008 BO${}_{16}$    & $-$0.0739 & 2019-08-18T08:36 & 295.6 &   $-$11.7 &   24.9 &     1   &   GYO1/9CU0.0 \\
2008 BO${}_{16}$    & $-$0.0620 & 2019-06-16T13:58 & 287.8 &   $-$11.1 &   24.8 &     1   &   SYO0/1CE0.0 \\
2008 ED${}_{69}$    & $-$0.0006 & 2019-07-09T04:58 & 247.5 &    63.0 &   22.9 &     1   &  GYO0/4CU12.7 \\
2013 CT${}_{36}$    & $-$0.0377 & 2019-10-12T01:55 & 261.1 &    21.9 &   12.5 &     8   &  GYO0/19CU0.0 \\
2013 CT${}_{36}$    & $-$0.0620 & 2019-06-16T13:58 & 287.8 &   $-$11.1 &   24.8 &     4   &   SYO0/1CE0.0 \\
2013 CT${}_{36}$    &  0.0190 & 2019-12-08T00:21 & 241.5 &    27.2 &   12.6 &     1   &   SYO0/1CE0.0 \\\botrule
\end{tabular}
\end{table}
\clearpage

\begin{table}
\caption{Summary of impacts on Jupiter\label{tab:Impacts}}
{\tabcolsep5.5pt%
\begin{tabular}{@{}ll@{}r}
\toprule%
Date & Mass (kg) & Reference\\
\hline
81-03-05         & 11 & \citet{Cook81}\\
94-07-16 &  $1.0 \times 10^{12}$  &\citet{Hammel95},\\
-- 24 &     &\citet{Harrington04}\\
09-07-19         &  $6.0 \times 10^{10}$  & \citet{SanchezLavega10}\\
10-06-03         &  $1.05$-$7.80$ $\times 10^3$            & \citet{Hueso10b,Hueso13}\\
10-08-20         &  $2.05$-$6.10$ $\times 10^3$    & \citet{Hueso13}\\
12-09-10         &  $5.00$-$9.50$ $\times 10^3$  & \citet{Hueso13}\\
16-03-17         &  $4.03$-$8.05$ $\times 10^3$  & \citet{Hueso18a}\\
17-05-26         &  $0.75$-$1.30$ $\times 10^3$  & \citet{Hueso18a}\\\botrule
\end{tabular}}
\end{table}
\clearpage

\clearpage

\clearpage

\clearpage

\clearpage

\clearpage

\clearpage

\end{document}